# Ablation threshold and temperature dependent thermal conductivity of high entropy carbide thin films

Milena Milich,[1*] Kathleen Quiambao-Tomko,[2] John Tomko,[1] Jon-Paul Maria,[4] Patrick Hopkins[1,2,3]

[1]Department of Mechanical and Aerospace Engineering, University of Virginia, Charlottesville, VA 22904, USA
[2]Department of Materials Science and Engineering, University of Virginia, Charlottesville, VA 22904, USA
[3]Department of Physics, University of Virginia, Charlottesville, VA 22904, USA
[4] Department of Materials Science and Engineering, Pennsylvania State University, University Park, PA 16802, USA

## ABSTRACT

High entropy carbides (HECs) are a promising new class of ultra-high temperature ceramics that could provide novel material solutions for leading edges of hypersonic vehicles, which can reach temperatures > 3,500 °C and experience extreme thermal gradients. Although the mechanical and thermal properties of HECs have been studied extensively at room temperature, few works have examined HEC properties at high temperatures or considered these materials' responses to thermal shock. In this work, we measure the thermal conductivity of a five-cation HEC up to 1200 °C. We find that thermal conductivity increases with temperature, consistent with trends demonstrated in single-metal carbides. We also measure thermal conductivity of an HEC deposited with varying $CH_4$ flow rate, and find that although thermal conductivity is reduced when carbon content surpasses stoichiometric concentrations, the films all exhibited the same temperature dependent trends regardless of carbon content. To compare the thermal shock resistance of HECs with a refractory carbide, we conduct pulsed laser ablation measurements to determine the fluence threshold the HECs can withstand before damaging. We find that this metric for the average bond strength trends with the theoretical hardness of the HECs as expected.

## 1 INTRODUCTION AND BACKGROUND

*E-mail: mm9bq@virginia.edu

The thermodynamic response of materials when subjected to immense thermal loads and shocks typical during hypersonic flight dictates the reliability and performance of hypersonic vehicles [1]. The leading edge of hypersonic vehicles is a region of intense thermal fluxes, the power densities of which can initiate failure of the leading edge. During hypersonic flight, the intense, localized and rapid heating of leading edges can routinely lead to several failure mechanisms of ultrahigh temperature materials. Thus, candidate materials for use in these extreme environments are usually carbon-based composites [2-10] and ultrahigh temperature ceramics (UHTCs) such as refractory borides and carbides due to their high melting points and high thermal conductivities [11, 12].

New classes of UHTCs have recently emerged based on the concept of entropy stabilization [13-23], where configurational entropy based on multiple refractory metal cations is used to stabilize the lattice. This has resulted in remarkable properties of UHTCs ranging, including ultralow thermal conductivities [14]. These reduced thermal conductivities of UHTCs as compared to their single-metal cation cousins are driven by the increased chemical disorder from the multiple different metal species that result in electron or phonon scattering events. However, unique to these multi-principal element solutions, local lattice distortions or chemical off-stoichiometries can additionally influence the thermal conduction processes [14, 24].

To exemplify the role that these defects specific to entropic stabilization in crystals can have on phonon conduction, consider that the thermal conductivity in non-metallic crystals is reduced as lattice "imperfections" reduce the phonon mean free path [25-28]. From the phonon's perspective, these imperfections can be in the form of difference in masses in a multi-element chemically disordered solid solution [29], such as in the entropy stabilized oxides. Figure 1a exemplifies this concept, which shows the thermal conductivity of various entropy stabilized oxides (ESOs) as a function of $6^{th}$ component mass from the ESOs studied by Braun *et al* [14] compared to the thermal conductivity of the base 5-cation oxide system with an average cation mass of ~54 amu. These ESOs are single-phase thin films, with rocksalt structures having a fixed oxygen anion sublattice; between each oxygen atom pair sits a cation randomly selected among the equiprobable distribution of five or six unique elements. These films included J14 ($Mg_xNi_xCu_xCo_xZn_xO$, x = 0.2), and five 6-cation oxides made up of the J14 composition plus an additional cation to include Sc (J30), Sb (J31), Sn (J34), Cr (J35), and Ge (J36). As shown in the experimental data in Figure 1a, the thermal conductivity of these ESO samples can exhibit values similar to an amorphous ESO, a-J14 (also shown in Fig. 1a, as reported by Braun *et al.* [14]), representing that these single crystalline ESO films possess "ultra-low" thermal conductivities (similar to amorphous phase). In particular, in the 6-cation oxides (J30 series), the thermal conductivity is reduced by nearly a factor of two relative to the J14 5-cation oxide. This reduction and ultra-low thermal conductivities of the J30 series is enabled by a local lattice distortion in the J30 relative to the J14 that occurs with the addition of the sixth cation.

The reduction in the phonon contribution to thermal conductivity in high entropy materials demonstrates immense promise to create materials with phonon thermal conductivities that can achieve "ultralow" values, even below those of the so-called "minimum limit" to thermal conductivity [30-33]. While this would be beneficial to heat shielding and insulations, this is detrimental to alleviate high heat fluxes and dissipation of localized hot spots that what would occur on a leading edge during hypersonic flight. For this, new classes of high entropy ceramics, such as high entropy carbides [23, 34, 35], can offer novel solutions.

Typical carbide structures can be viewed as an arrangement of interstitial carbon in a simple metallic lattice. In many cases, carbides take on the cubic rocksalt or hexagonal structures. Deviations from ideal stoichiometry are quite common, and specific structural phases exist within a broad range of carbon content. As such, point defects, particularly carbon vacancies and interstitials can have a dramatic effect on material properties [36-40]. An important question to consider is how thermal properties of high entropy carbide systems respond to deviations in carbon content. While the thermal properties of refractory ceramics are generally well-studied, explorations of high entropy carbide systems remain few and far between, especially with regard to carbon stoichiometry and defects.

In a prior work, we measured the thermal conductivity of novel high entropy carbide (HEC) thin films with nominal compositions $(HfZrTaMoW)_{0.2}C_{1-x}$ and $(HfNbTaTiZr)_{0.2}C_{1-x}$, and their dependence on carbon stoichiometry, $x$ [23]. These HECs transition from electrically-conducting, with electrons dominating the thermal transport in a material with primarily metallic bonding, to ceramic-like systems with primarily covalent bonding, where thermal conductivities are largely dominated by the phononic subsystem, as shown in Figure 1b. While the electronic contribution to thermal conductivity remains at a constant when the crystal is primarily covalently bonded, a combination of changes in film morphology, point defect scattering, and phase precipitation systematically lower the phononic contribution to thermal conductivity with further increased carbon content in the films.

These prior results were reported at room temperature, and thus to further understand and evaluate the utility of HEC materials, in this work we study the temperature dependent thermal conductivity of a series of HEC films from room temperature to 1,200 °C to assess the role that carbon stoichiometry has on the thermal conduction processes at elevated temperatures. Refractory carbides are well known to have an increasing thermal conductivity with increasing temperature due to carbon vacancies causing a high residual electrical resistivity. We find a similar trend in the HEC films studied in this work.

To additionally evaluate the performance of these HEC films under the extreme heat fluxes typical of hypersonic vehicle leading edges, we study the damage induced in HEC films by an ultra-fast laser pulse of varying power. By measuring the ablated area and corresponding laser fluence, we can calculate the damage fluence threshold of the material - the minimum power density at which damage will occur. This value is effectively a measure of the average bond strength in the material, as the time scale of ablation (and the laser pulse) is significantly faster than that of conduction, and so little of the pulse energy is dissipated as heat via conduction. We find that the damage threshold varies only slightly between HECs of different metal compositions with equal carbon content. Furthermore, we observe that the HECs' damage thresholds trend linearly with theoretical hardness, consistent with the fact that both are measures of bond strength.

## 2 EXPERIMENTAL DETAILS

### 2.1 Film growth – High Power Impulse Magnetron Sputtering (HiPIMS)

In this work we consider three HEC compositions: HEC-3 (HfNbTiTaZrC), HEC-4 (HfNbTiTaVC), and HEC-9 (HfNbTiTaWC). The films studied in this work are similar to those reported in our prior work, deposited with reactive bipolar high power impulse magnetron sputtering (HiPIMS) on c-plane sapphire [35]. This technique allows greater control over both stoichiometry and microstructure, and ensures a more uniform thin film composition. With conventional direct current and radio frequency magnetron sputtering techniques, it has been shown that carbon will precipitate to enclose the metal carbide grains before stoichiometric ratio has been reached, leaving carbon deficient grains surrounded by graphitic carbon. HiPIMS allows for the deposition of stoichiometric metal carbides by controlling the deposition parameters, leading to improved mechanical, thermal, and chemical properties. The deposition technique and film characterization of two HEC thin films are further described in our prior works [23, 35].

### 2.2 Thermal conductivity measurements – Time domain thermoreflectance

The thermal conductivity of the HEC films was measured using time-domain thermoreflectance (TDTR) [41-43]. Briefly, the output from a pulsed, 800 nm wavelength Ti:Sapphire laser system (Spectra Physics Tsunami) at a frequency of 80 MHz and bandwidth of 10.5 nm, is split into the pump and probe paths. The pump path passes through an electro-optical modulator (EOM). These pump pulses are then focused on the sample surface using a 10X infinity corrected microscope objective creating a modulated heating event on the sample surface. The probe path is passed through a mechanical delay stage, incrementally varying the arrival time between the pump and probe, and then focused into the middle of the pump spot on the sample surface. The back reflection of the probe is directed to a Si photodiode where the in-phase and out-of-phase voltages are recorded through a lock-in amplifier, monitoring changes with respect to the modulated heating event from the pump. We monitor the ratio of the in-phase to out-of-phase lock-in response and relate this ratio to the complex temperature change on the surface of a multilayer solid subjected to a frequency modulated train of delta functions as a thermal source to the heat equation in cylindrical coordinates, as described previously. We measure the thermal conductivity of the HEC thin films from room temperature to 1,200 °C. We control the temperature of our sample via resistively heated Linkam TS1500 under vacuum conditions.

In our previous work, we only studied the thermal conductivity at room temperature, and thus we used an 80 nm Al film as an opto-thermal transducer, as necessitated by TDTR. However, due to the high temperatures studied in this work, we use a thin W film instead of the Al film, enabling higher temperature measurements. The room temperature thermal conductivity of a series of these W coated HEC films is measured and compared to the thermal conductivity of a series of TiC films grown under similar conditions to those of the HEC films, shown in Figure 2. We show similar trends to the thermal conductivities reported in our prior work for the HEC films we study in this work. The TiC, however, is observed to have a very consistent thermal conductivity for $CH_4$ flow rates beyond 0.5 sccm. We approximate the $CH_4$ flow rate at which stoichiometric carbon content is reached to be where this leveling occurs, and electronic transport no longer dominates the total thermal conductivity. The wide ceramic transition zone observed for TiC, or range of $CH_4$ flow rates in which stoichiometric TiC is achieved, is attributed to its low valence electron concentration (VEC) of 8, and thus its high carbon affinity compared to the higher VEC HEC films. It is important to note that the reduction in thermal conductivity vs. $CH_4$ flow rate that we observe in the HiPIMS-deposited HEC films is this work occurs at lower $CH_4$ flow rate than in our prior work. This is due to the fact that in our prior work, the HEC films were deposited via conventional magnetron sputtering, where in this work, HiPIMS deposition was used.

### *2.3 Ablation threshold measurements*

We conducted ablation-based measurements on three HEC thin films to determine the threshold fluence at which damage occurs. A pulsed 1040 nm laser (Spectra Physics Spirit) is frequency doubled to 520 nm by a Hiro Light Conversion second harmonic generator. The 520 nm beam is then directed through a half-wave plate and beam splitter for power control, and is focused by a 10X infinity-corrected microscope objective to a 17 μm $1/e^2$ diameter spot size. A camera is used to view the damage of the sample surface. Each measurement location on the sample was damaged by a single ~200 fs pulse with instantaneous incident laser fluences ranging from 3.3 – 9.1 $TW/cm^2$. The average power of the pulsed laser at the sample surface was recorded for each fluence level. The resulting damage areas were imaged with SEM, and imageJ was used to measure the diameter of damage from the micrographs. The damage area follows a logarithmic trend, with a sharp increase in area between the lower fluences, as can be seen in Figure 3. The damage area then approaches a saturation point at higher fluences, where the damaged region no longer increases in area as the laser fluence is increased, matching the expected logarithmic trend. We calculate the damage fluence threshold by fitting the damage area vs fluence data to the following equation: $D^2 = 2\omega_0^2 \ln\left(\frac{F}{F_{th}}\right)$, where D is the diameter of the ablation area, $\omega_0$ is the beam waist, F is the incident laser fluence, and $F_{th}$ is the fluence threshold [44].

## 3 RESULTS AND DISCUSSION

Temperature dependence of thermal conductivity for thin films with varying carbon content is shown in Figure 4. For these measurements, we took TDTR data while the films were heated by a temperature stage under vacuum conditions. We observed a lower thermal conductivity in films with higher carbon content at all temperatures. This trend is due to the lowered electron contribution to thermal conductivity when more of the bonds are covalent. The more carbon that is present in the film, the more bonds are covalent, and phonon contribution to thermal conductivity increases relative to that of the electronic contribution [23]. Additionally, phonon scattering due to interstitial carbon atoms and graphitic carbon precipitates may play a role in reducing thermal conductivity when carbon content surpasses stoichiometric concentrations. Similarly, the observed increase in thermal conductivity with temperature consistent across all methane flow rates, is due to the presence of the metallic bonding and carbon defects. At higher temperatures, there is a greater concentration of free electrons, and thus an increased electron contribution to the total thermal conductivity. However, increased temperatures also result in higher phonon-defect scattering rates. The observed increase in thermal conductivity with temperature is thus the superposition of these two effects, with the increase in carrier concentration being the dominating factor. This trend is in agreement with trends previously shown in metal carbides [36, 39].

We also measure the temperature dependent thermal conductivity of a similarly deposited TiC thin film from room temperature to 1200°C and compare it to a stoichiometric HEC-3 thin film, shown in Figure 5. The room temperature conductivities of both films are significantly lower than previously reported for bulk TiC [36], due to the additional scattering caused by the smaller grain sizes and higher defect concentrations driven by carbon off stoichiometry [23, 45]. Additionally, as previously observed, the HEC thin film has a lower thermal conductivity than the TiC film due to scattering caused by the additional masses present. With regards to the temperature dependence of thermal conductivity in the sputtered thin films, the TiC follows the same nonlinear increasing trend as the HEC, and again consistent with the observed temperature dependence in thermal conductivity of metal carbides [36].

To study the resistance of the HECs to thermal shock, we measure the fluence threshold of three five-cation HEC thin films deposited under the same $CH_4$ flow rate and again compare the results to that of a similarly-deposited TiC film. At each damage location, the sample is subjected to a ~200 fs, 520 nm laser pulse with a fluence ranging from 3.3 – 9.1 $TW/cm^2$, as previously mentioned. The samples are imaged at 250X magnification with SEM to obtain precise measurements of damage diameter, which are then fitted to the logarithmic model presented earlier in the paper to determine the fluence threshold. The time scale of the laser pulses is sufficiently short such that negligible energy is dissipated via thermal conduction, and the area of damage is considered a direct indicator of the amount of energy it takes to break bonds and eject material from the sample surface. To ensure that the HEC and TiC films have similar absorptions at the wavelength used for ablation, we measure the reflectivity of each sample and find that the samples all fall in the range of 30-43% power reflected.

We observe slight differences between the fluence threshold of the samples, that appear to trend with theoretical hardness, as calculated in Hossain *et al.* [46], shown in Figure 6. This is expected, as both bond stiffness and fluence threshold are indicators of bond energy, and thus can also be predicted by the VEC. The VEC determines the average carbon affinity of the transition metals and therefore the bond strength between each metal and carbon in the lattice. We see that HEC-

3, which has a VEC of 8.4, has a higher fluence threshold and thus higher bond energy than HEC-4, which has a VEC of 8.6, or HEC-9 with a VEC of 8.8, because it takes a greater amount of energy to break ZrC bonds than VC or WC. Additionally, we notice that although the fluence thresholds are comparable between the different HECs, we observe significant differences in the magnitude of damage between the samples. This appears to trend roughly with the lattice parameter of the 5th cation added to the HEC. The HEC-3 sample, which has Zr as its 5$^{th}$ cation, experienced less damage than the other HECs at the same laser fluence. ZrC also has a larger lattice parameter than VC or WC, and this appears to yield a greater tolerance to thermomechanical shock.

## 4 SUMMARY

In this work we measure the temperature dependent thermal conductivity of $(HfNbTaTiZr)C_{1-x}$ deposited by HiPIMS under three different $CH_4$ flow rates up to 1200 °C. We find that at all temperatures, HECs with higher carbon content have a lower thermal conductivity due to the shift towards majority covalent bonding from metallic, and an increase in phonon scattering caused by interstitial carbon. Although HECs with greater than stoichiometric carbon content nominally exhibit lowered thermal conductivities, all of the HEC films studied exhibited the same increase in thermal conductivity with temperature. We also compare a stoichiometric HEC film with a similarly deposited thin film refractory carbide, TiC, and find that the HEC again demonstrates the same increase in thermal conductivity with temperature as the single-metal carbide, but with a nominally lower thermal conductivity due to additional scattering from the 5 metal cations.

In studying the thermal shock resistance of three HEC films, we find that the fluence threshold at which damage occurs is comparable, and somewhat higher, than that of TiC. Amongst the HECs, the damage threshold appears to trend with the theoretical hardness of the material, with harder materials demonstrating greater resistance to damage from ablation. This is as expected because both fluence threshold and hardness are indicators of average bond strength. Additionally, we observe that although the calculated fluence thresholds are largely comparable between the HECs, the extent of damage at each fluence level varies significantly, and the damage area measured at the highest fluence tested appears to trend with average lattice parameter. The least damaged sample was that with the largest average distance between atoms.

**ACKNOWLEDGEMENTS**

We appreciate support from the Office of Naval Research, Grant Number N00014-22-1-2139.

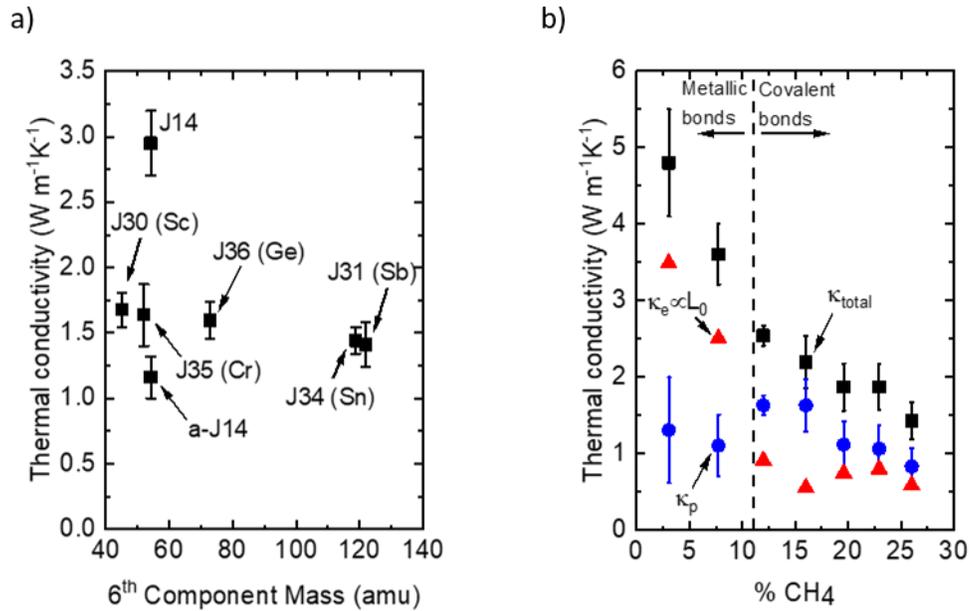

Figure 1. (a) Thermal conductivity of entropy stabilized oxide (ESO) thin films as a function of 6th component mass. The reduction in thermal conductivity from the 5 cation ESO (J14) to the various 6 cation ESOs (J30-series) is due to the additional lattice distortion that the 6th cation introduces to the lattice, thus offering additional phonon scattering mechanisms, and resulting in thermal conductivity near that of the amorphous 5-cation oxide (a-J14). These results are detailed in Bruan *et al* [14]. Sample key: J14 = $(MgNiCuCoZn)_{0.2}O$; J30 = J14+Sc; J31 = J14+Sb; J34 = J14+Sn; J35 = J14+Cr; J36 = J14+Ge. (b) Thermal conductivity of $(HfZrTaMoW)_{0.2}C_{1-x}$ high entropy carbide (HEC) thin films as a function of methane flow rate (% $CH_4$) in the plasma. A high $CH_4$ content results in a change in bond type from metallic to covalent, along with an increased precipitation of excess carbon. This inversely impacts the electronic contribution to thermal conductivity, with a continued decrease in electron thermal conductivity as a function of increase % $CH_4$. These results are detailed in Rost *et al* [23].

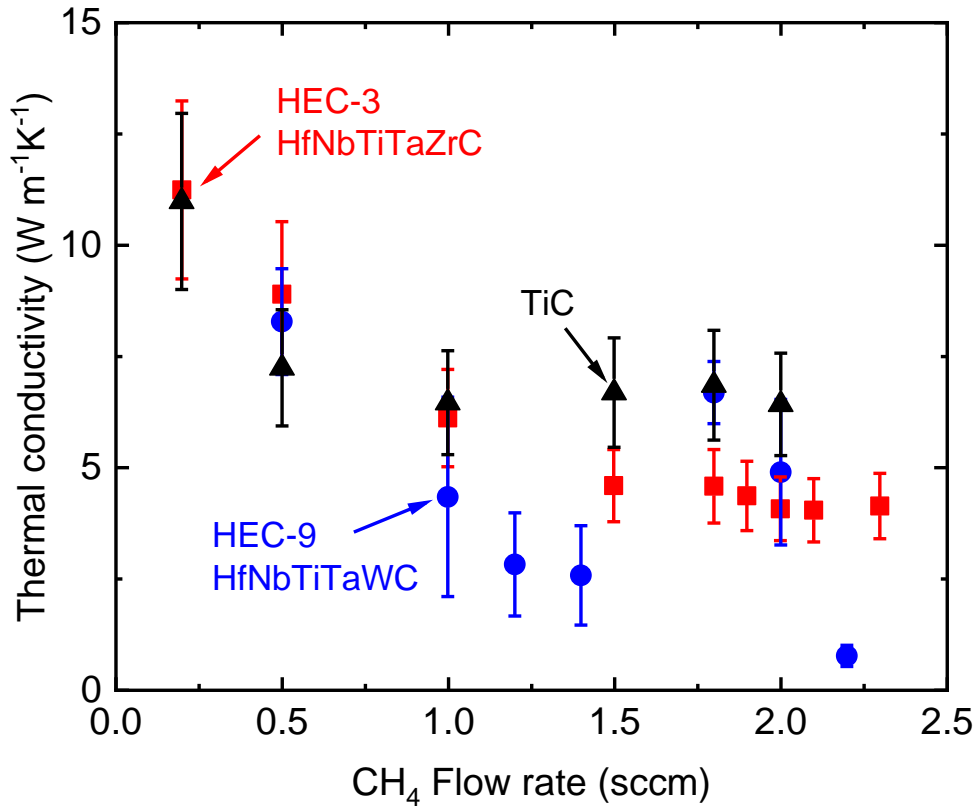

Figure 2. Room temperature thermal conductivity of HEC thin films as a function of $CH_4$ flow rate. At low $CH_4$ flow rate, the films are electronically conductive, while at high $CH_4$ content, there are more covalent bonds and thermal conductivity is phonon-driven, leading to a lowered total thermal conductivity. Furthermore, interstitial carbon and carbon precipitates create additional phonon scattering events as carbon content exceeds stoichiometric concentration.

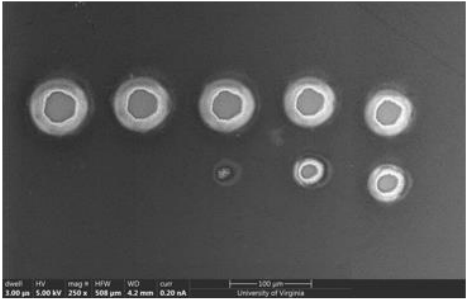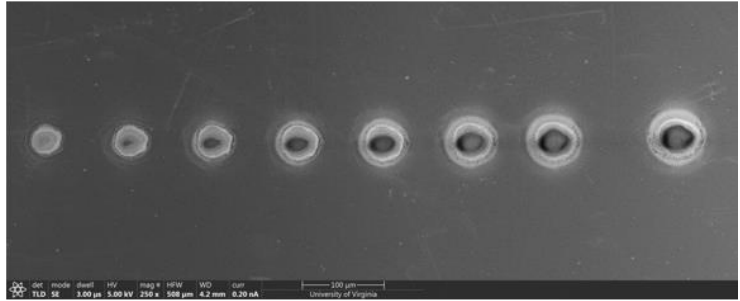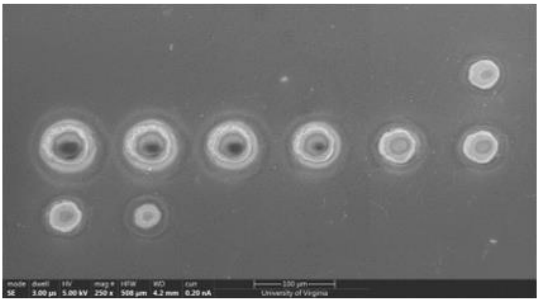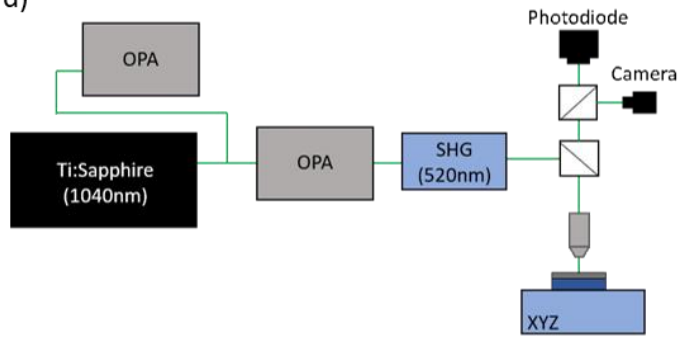

Figure 3. SEM micrographs of ablation damage for varying fluences on (a) HEC-3 (HfNbTiTaZrC), (b) HEC-4 (HfNbTiTaVC), and (c) HEC-9 (HfNbTiTaWC). The damage diameter used to calculate fluence threshold is the outer diameter of the hole, the innermost white ring observed in the SEM micrographs. The experimental setup is shown in (d).

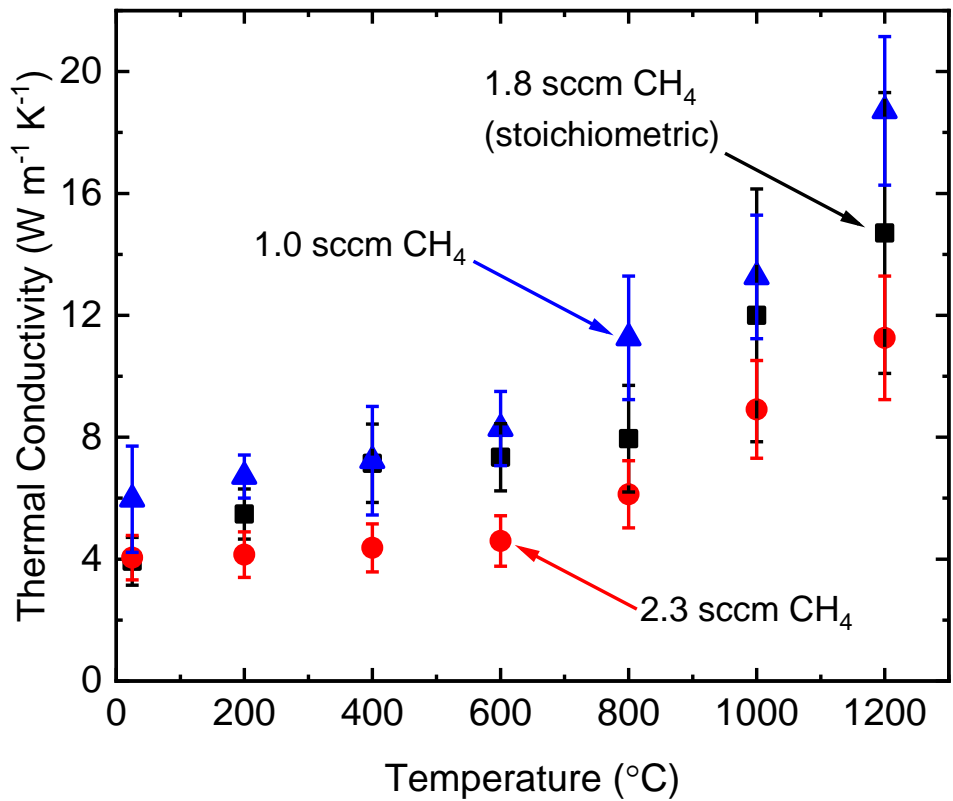

Figure 4. Temperature dependence of thermal conductivity of HEC-3 thin film with varying $CH_4$ flow rates. At higher temperatures, there is a greater concentration of free electrons, and thus an increased electron contribution to the total thermal conductivity. This increase in carrier concentration with increased temperature tempered by increased phonon-defect scattering rates leads to the observed increase in thermal conductivity with temperature, in agreement with trends observed in metal carbides [36, 39].

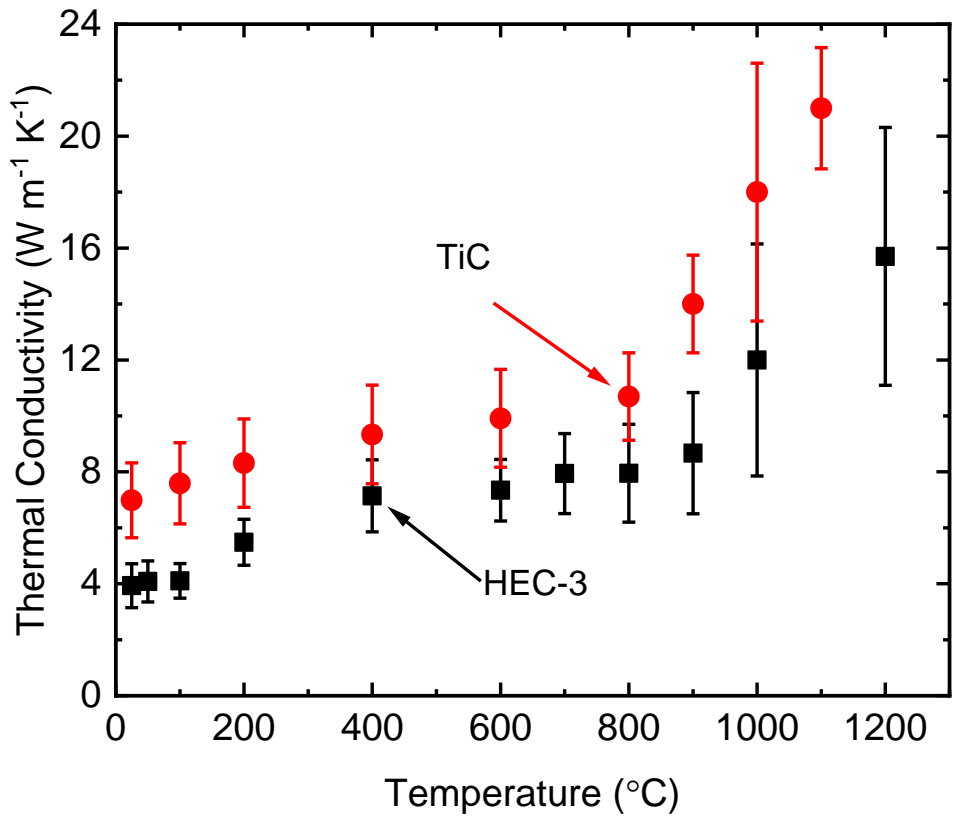

Figure 5. Temperature dependence of HEC-3 thermal conductivity in comparison to TiC thin film.

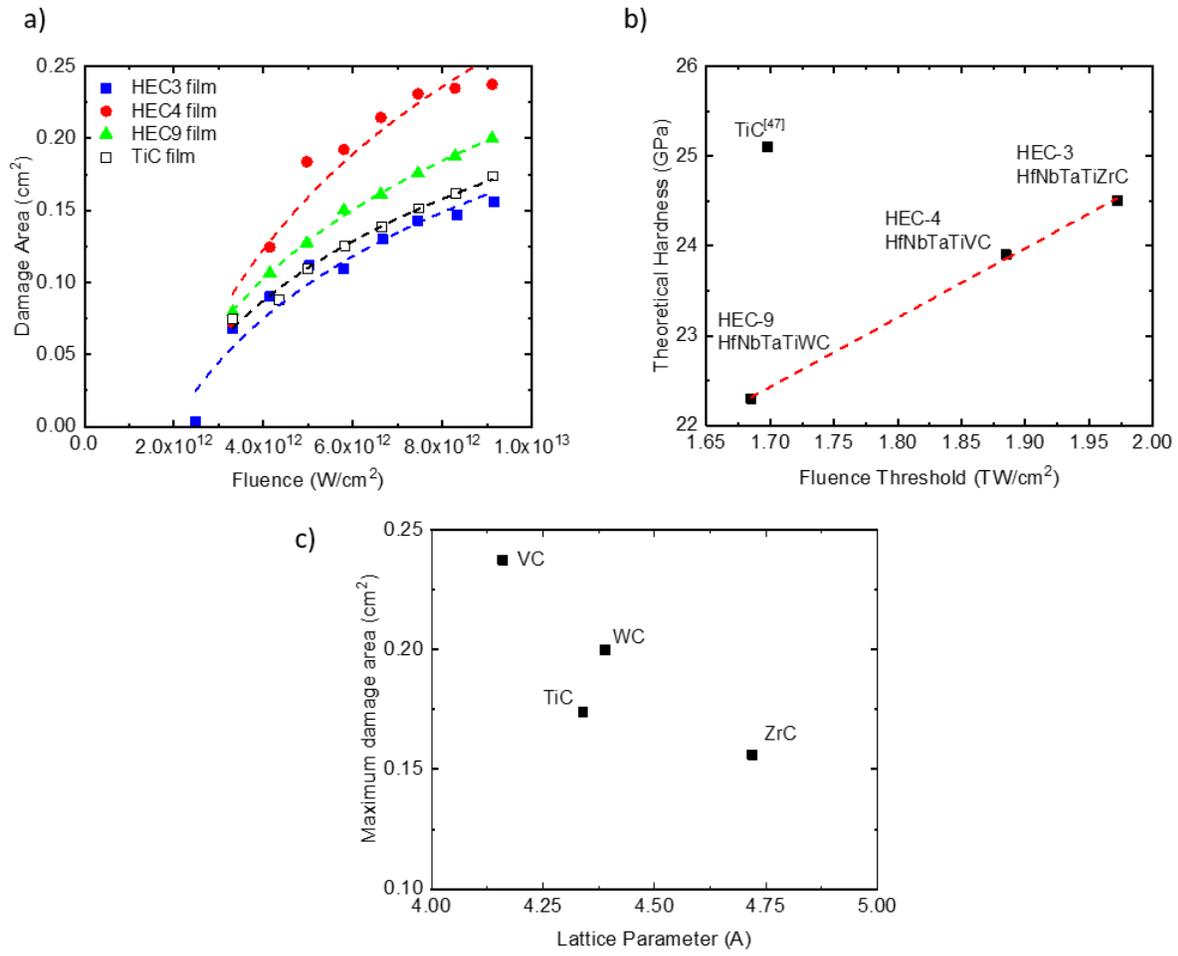

Figure 6. (a) Damage area as a function of laser fluence for three HEC thin films. Dashed lines represent the fitted logarithmic model used to calculate fluence threshold. (b) Fluence thresholds vs. theoretical hardness. Both fluence threshold and hardness are indicators of bond strength, and we observe a near linear agreement between the two. (c) Lattice parameter of the single-component carbide corresponding to the 5th cation added, and the area damaged at 9 TW/cm$^2$.